\documentclass[journal]{IEEEtran}
\usepackage{enumerate}
\usepackage{stfloats}
\usepackage{graphicx}
\usepackage{hyperref}
\usepackage{multirow}
\usepackage{array}
\usepackage{booktabs}
\usepackage{amsmath}
\usepackage[english]{babel} 
\usepackage{float}

\makeatletter
\usepackage{url}
\g@addto@macro{\UrlBreaks}{\UrlOrds}
\let\runtitle\@title
\makeatother
\usepackage{xcolor}
\usepackage{amssymb}
\usepackage{gensymb}
\usepackage{comment}
\usepackage{graphicx}
\usepackage[utf8]{inputenc}
\usepackage{caption}
\usepackage[margin=2cm]{geometry}
\usepackage{array}
\usepackage[utf8]{inputenc}

\usepackage[
backend=biber,
sorting=none
]{biblatex}
\begin{document}

\title{Shap-MeD}

\makeatletter
\let\Title\@title
\makeatother

\author{
  \IEEEauthorblockN{Nicolas Laverde \IEEEauthorrefmark{1}}, \IEEEauthorblockN{Melissa Robles \IEEEauthorrefmark{2}},   \IEEEauthorblockN{Johan Rodríguez \IEEEauthorrefmark{3}}\\
\IEEEauthorblockA{ Universidad de los Andes.
Bogotá, Colombia\\
E-mail: \IEEEauthorrefmark{1}n.laverdem@uniandes.edu.co, \IEEEauthorrefmark{2}mv.robles@uniandes.edu.co y \IEEEauthorrefmark{3}jd.rodriguezp1234@uniandes.edu.co 
}
}

\markboth{\Title}
{\Title}

\maketitle

\begin{abstract}
We present Shap-MeD, a text-to-3D object generative model specialized in the biomedical domain. The objective of this study is to develop an assistant that facilitates the 3D modeling of medical objects, thereby reducing development time. 3D modeling in medicine has various applications, including surgical procedure simulation and planning, the design of personalized prosthetic implants, medical education, the creation of anatomical models, and the development of research prototypes. To achieve this, we leverage Shap-e, an open-source text-to-3D generative model developed by OpenAI, and fine-tune it using a dataset of biomedical objects. Our model achieved a mean squared error (MSE) of 0.089 in latent generation on the evaluation set, compared to Shap-e's MSE of 0.147. Additionally, we conducted a qualitative evaluation, comparing our model with others in the generation of biomedical objects. Our results indicate that Shap-MeD demonstrates higher structural accuracy in biomedical object generation.
\end{abstract}

\section{Introduction}
The new trend in healthcare is personalized treatments, moving away from universal approaches and \textit{one-size-fits-all} treatments. According to the National Health Service (NHS) of England, universal treatments exhibit a 70\% inefficiency rate among patients, highlighting the need for personalized treatments \cite{TP}. However, implementing a personalized treatment approach is unfeasible with traditional manufacturing processes, as it involves time- and labor-intensive tasks. In contrast, 3D-printed products have been well received by patients \cite{GOYANES201771}. This trend underscores the necessity for the industry to transition toward new technologies that enable personalization.\\

\begin{figure}[H]
    \centering
    \includegraphics[width = 0.39\textwidth]{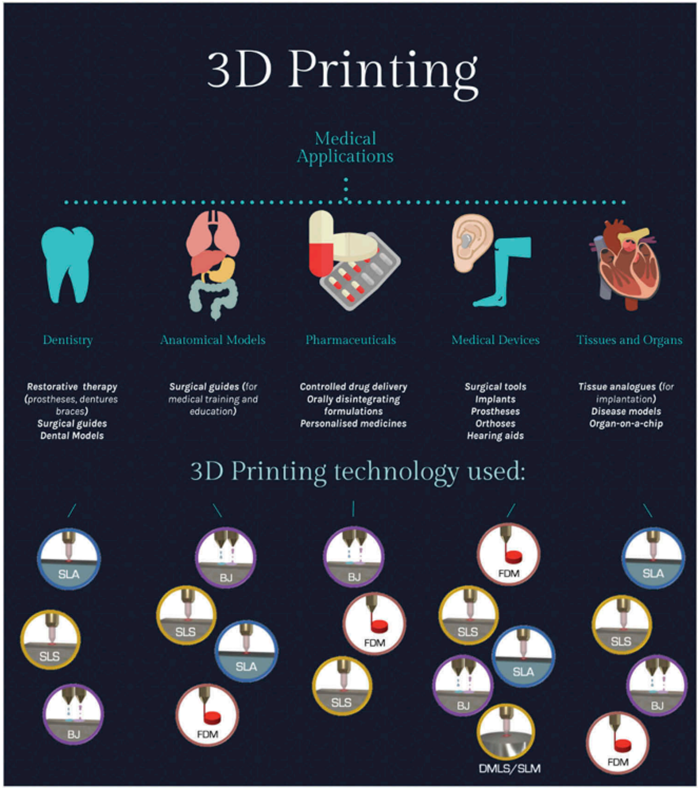}
    \caption{Current applications of 3D printing in medicine and healthcare. \cite{img}}
    \label{fig:aplicaciones}
\end{figure}

3D printing was introduced more than three decades ago. Since then, this technology has radically transformed the manufacturing industry. The original purpose of this technology was the creation of engineering prototypes, including automotive parts, fashion accessories, and even architectural models, which were produced using this method. In fact, it is still used for these purposes today \cite{barnatt20133d, Chowdhry2013, Ventola2014}. However, its applications extend far beyond these initial uses. 3D printing has undoubtedly revolutionized the healthcare system, as it enables the production of customized objects, making it particularly suitable for the personalization of prosthetics, implants, drug delivery systems, tissues, and medical devices \cite{Sahlgren2017}. Figure \ref{fig:aplicaciones} illustrates the five major categories in which 3D printing is employed.\\

One of the most pressing issues in healthcare that 3D printing addresses is the shortage of organs for transplantation. According to organ donation statistics, approximately 20 patients die each day while waiting for a transplant \cite{Shinde2022}. In addition to the limited availability of organ donors, another significant challenge is ensuring the biocompatibility of the donated organ \cite{JiménezOliver_2023}. 3D printing has the potential to generate tissues with patient-specific dimensions and materials, which could significantly contribute to organ transplantation efforts \cite{zheng20193d}.

Significant advancements have been made in 3D printing for personalized treatments. Structures can be generated from a digital 3D file using \textit{computer-aided design} (CAD) software and computer vision techniques, including magnetic resonance imaging (MRI) or 3D tomography \cite{Trenfield2018}. Although the results are highly promising, it is important to note that the level of personalization achievable with computer vision techniques is limited by the quality of the acquired images.\\

The objective of this project is to develop an assistant dedicated to the creation of anatomical models and organs, conditioned by text or images. This aims to facilitate the development of research prototypes for tissue engineering and explore their potential applicability in prosthetic transplants. Additionally, the system is envisioned as an educational tool, benefiting students by enhancing their learning and understanding of anatomy and medicine. In line with the concept of treatment personalization, we aim to incorporate patient-specific information into the design of these 3D representations. This will be achieved through \textit{fine-tuning} generative models based on diffusion models using biomedical objects. While strategies for generating conditioned 3D meshes will be presented later, it is important to emphasize that none of the existing approaches focus specifically on the biomedical domain.

\section{Background}
Determining the methodology for representing three-dimensional objects presents a challenge due to the inherent complexity of shapes, variations in lighting, and spatial dimensions. Various strategies are employed to address different aspects of 3D representation, including the use of polygonal meshes, point clouds, and implicit function-based approaches. Examples of these techniques include the STL, XYZ, and NeRF formats, which correspond to each of the aforementioned methods, respectively. This section describes the different representations necessary to understand the models used throughout the project.

\subsection{STL}
The three-dimensional representation of an object using an STL file involves a geometric description through polygonal meshes. This file format encodes the object's surface as a network of triangles, where each vertex defines a point in three-dimensional space. An STL representation is thus a matrix of dimensions $(T,3,3)$, where $T$ represents the number of triangles in the model, and each triangle is represented by a $3\times3$ matrix of the form

\begin{align*}
\begin{bmatrix}
    x_1 & y_1 & z_1 \\
    x_2 & y_2 & z_2 \\
    x_3 & y_3 & z_3 \\
\end{bmatrix}.
\end{align*}
Additionally, the components of the unit normal vector to each triangle are included, ensuring that it points outward according to the model's orientation. This representation enables the transfer of 3D models between different programs and systems, facilitating their use in applications such as design, engineering, and manufacturing.

\subsection{OBJ}
The OBJ format, also known as the Wavefront OBJ file, is a simple representation for 3D objects, originally developed for the visualizer The Advanced Visualizer by Wavefront Technologies \cite{Wavefront}. This format consists of a plain text file that summarizes the object's information, including texture and color details. Due to these features, the OBJ format has gained greater popularity in 3D printing compared to the STL format.\\

This format allows for the representation of more complex surfaces beyond a simple arrangement of triangles. A set $V$ of vertices $v_1, v_2, \dots, v_n \in \mathbb{R}^3$ is defined, and subsets of these vertices are specified to form the corresponding surfaces or hyperplanes. Additionally, the text file includes information related to the textures and colors of the defined surfaces.
\\

Unlike the STL format, this file type not only significantly enhances 3D representation by offering greater flexibility in defining hyperplanes but also enables the inclusion of \textit{Freeform} curves and surfaces. These curves are constructed using polynomial interpolation between a finite set of vertices, resulting in the approximation of smooth and continuous 3D surfaces that conform to the selected vertices, thereby generating a more seamless structure.


\subsection{NeRF}
\label{subsec:nerf}
The NeRF 3D representation was proposed in 2022 \cite{NeRF} as an alternative to the classical representations used until then. It is based on the construction of an implicit function that maps tuples $(\mathbf{x}, \mathbf{d})$ of coordinates and directions to tuples $(\mathbf{c}, \sigma)$ of colors and densities. The input coordinates, denoted as $\mathbf{x} = (x, y, z) \in \mathbb{R}^3$, are represented as points, while the directions are expressed using two angles, $(\theta, \phi) \in \mathbb{R}^2$, which define the viewpoint from which the object is observed. On the other hand, colors follow the RGB representation $(c_R, c_G, c_B)$, and densities are real numbers that symbolize opacity or the \textit{amount of matter} at a given location within the three-dimensional volume of the scene. In summary, the function can be characterized as  

\begin{align*}
    F_{\Theta}: \mathbb{R}^5 &\rightarrow \mathbb{R}^4\\
    (\textbf{x}, \textbf{d}) &\mapsto (\textbf{c}, \sigma).
\end{align*}  

The function is parameterized using a Multi-Layer Perceptron (MLP) with weights $\Theta$, which gives $F_\Theta$ the name of an implicit function. Given this function, an image can be generated for a specific viewpoint by assigning to each pixel the color obtained from the neural network.

\subsection{STF}
\label{subsec:stf}

STF representation (Signed Distance Functions and Texture Fields) refers to the formulation of an implicit function as a method for three-dimensional representation, characterized by generating signed distances or textured colors as outputs. The SDF representations (Signed Distance Functions) are defined through a function

\begin{align*}
    G_{\Theta}:  \mathbb{R}^3 &\rightarrow \mathbb{R}\\
    (x,y,z) &\mapsto d.
\end{align*}

where $|d|$ represents the distance from the point $(x,y,z)$ to the closest point on the object's surface, and the sign $\operatorname{sign}(d)$ is negative if the point is inside the object and positive if it is outside. This function is considered implicit since the object's surface is defined as the set of coordinates $(x,y,z)$ that satisfy $G_\Theta(x,y,z) = 0$.
This SDF representation has been employed in generative models such as DMTet \cite{dmtet}. Other approaches extend this concept by incorporating additional information beyond signed distance, integrating texture information (Texture Fields) and RGB color data.

\section{Related Work}\label{sec:relatedwork}

\subsection{Point-E}\label{sec:pointe}
At the time of publishing Point-E \cite{Point-E} in 2022, all state-of-the-art models required extensive inference time on multiple GPUs to generate just a single example, making it impractical to democratize 3D content generation from text for the general public. Therefore, this work proposes an alternative approach that generates a 3D point cloud from text. A point cloud is defined as an array of tuples, where each tuple represents a point containing three entries for the three-dimensional coordinates and three additional entries for the RGB color channels.
\\

To generate point clouds like the one previously described, the architecture is based on transformers \cite{vaswani_attention_2017} and Gaussian diffusion models \cite{ho2020denoising}. These models rely on a stepwise noise addition process, defined as
\begin{align*}
    q(x_t|x_{t-1}): \mathcal{N}(x_t; \sqrt{1-\beta}x_{t-1}, \beta_t\textbf{I})
\end{align*}

Where \( x_{t-1} \) and \( x_t \) represent two consecutive steps in the noise addition process applied to an example \( x \)—in this case, a point cloud—the generative process is modeled as a normal distribution conditioned on the noisy example from the previous step, \( x_{t-1} \). Additionally, a noise scheduling variable, \( \beta \), gradually adjusts the noise level at each step, ensuring controlled variance to prevent excessive fluctuations between iterations. This noise addition process is approximated using a transformer architecture \cite{vaswani_attention_2017}, which models the inverse conditional probability \( p_\theta(x_{t-1} \mid x_t) \) by predicting the noise component \( \epsilon \). \\

\begin{figure}[H]
    \centering
    \includegraphics[width = 0.40\textwidth]{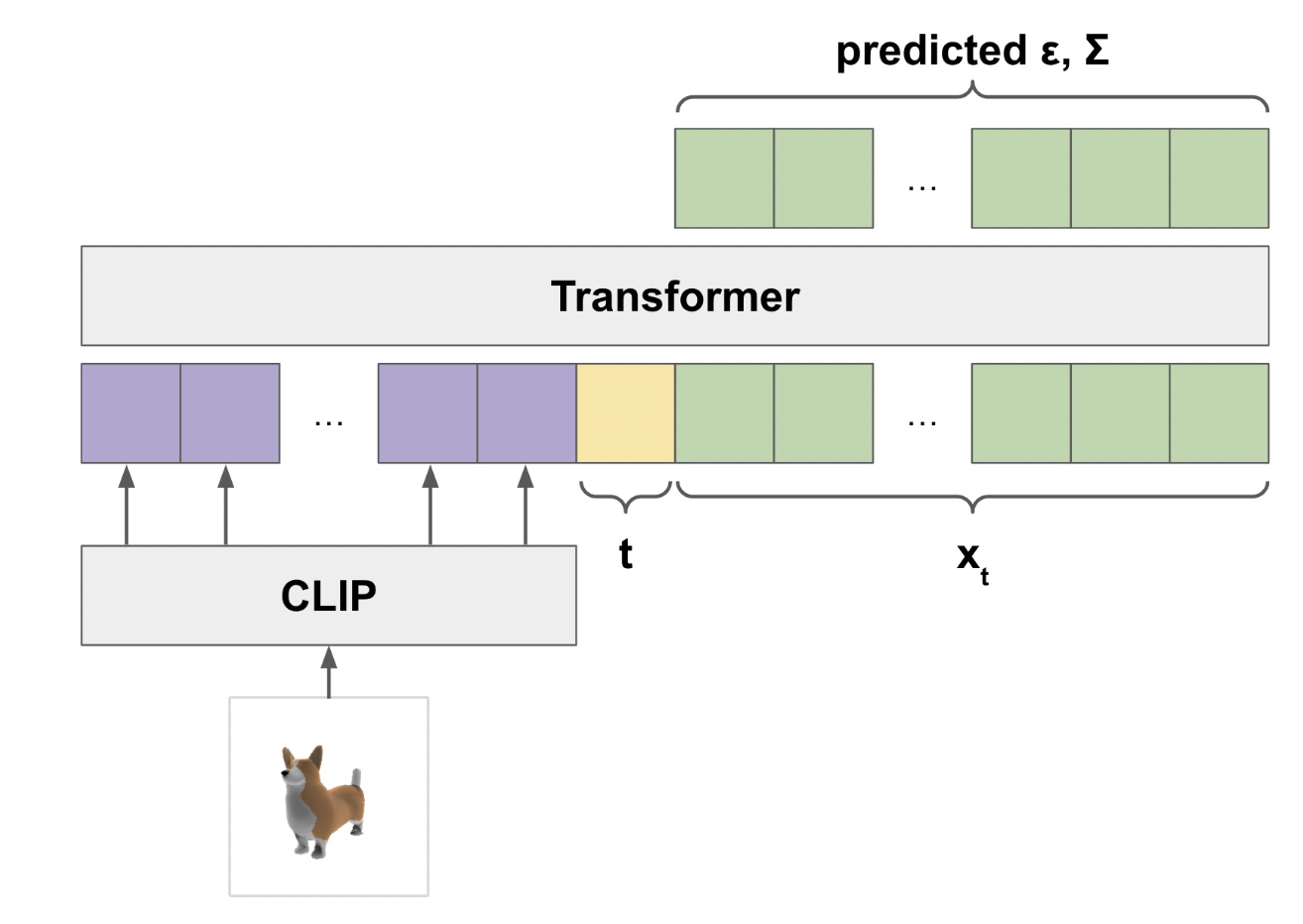}
    \caption{Transformer-Based Diffusion Architecture for Generating a Semantically Coherent Point Cloud from an Image and a Normally Distributed Noisy Point Cloud \cite{Point-E}}
    \label{fig:point-e-difussion}
\end{figure}

While this approach is effective for generating meaningful point clouds from randomly initialized, normally distributed noise, modifications are necessary to enable generation based on a given textual description. Typically, a conditioning process is employed, where the noisy point cloud is concatenated with a textual representation as input to the transformer architecture, allowing for text-driven inference \cite{oord2018neural}. However, empirical findings during the development of Point-E \cite{Point-E} revealed that conditioning on images rather than text yielded superior results. Consequently, the final architecture utilizes a diffusion model based on transformers, conditioned on an image representation obtained from a ViT-L/14 CLIP model \cite{radford2021learning} and a noisy point cloud. This setup enables the prediction of a semantically meaningful point cloud that accurately represents the content of the given image, as illustrated in Figure \ref{fig:point-e-difussion}.

As observed, this architecture solely enables the generation of point clouds from images, necessitating the introduction of a component that generates images from text. This component is a GLIDE model \cite{nichol2022glide}, a text-conditioned diffusion model that, unlike CLIP \cite{radford2021learning}, does not require labeled classification text to generate images. To optimize the use of GLIDE, fine-tuning was performed using a proprietary dataset comprising millions of images, each created from 3D renders using Blender \cite{blender}. Specifically, 20 images per model were generated, each corresponding to a randomly chosen camera angle.
\\

This architecture consists of three sequential generative components:
\begin{enumerate}
\item \textbf{Text-to-Image Generator}: A GLIDE architecture is employed to convert textual descriptions into images.
\item \textbf{Low-Resolution Point Cloud Generator}: A transformer-based architecture, as shown in Figure \ref{fig:point-e-difussion}, is used to generate a low-resolution point cloud consisting of 1,000 points from the image produced in the previous step.
\item \textbf{High-Resolution Point Cloud Generator}: Similar to the previous step, a transformer-based architecture (also illustrated in Figure \ref{fig:point-e-difussion}) is employed. However, in this case, the architecture is conditioned not only on the image from step 1 but also on the low-resolution point cloud generated in step 2. Specifically, a portion of the input point cloud is replaced with the generated low-resolution point cloud to produce a high-resolution point cloud containing 4,000 points. \\
\end{enumerate}

In the architecture described above, the training of the first step relies on text-image tuples, where synthetic views are created using a dataset of 3D objects. To complete the training process for the entire pipeline, high-resolution point clouds are generated from the 20 available views per object. A sampling process is then applied to obtain point clouds containing 4,000 and 1,000 points, respectively. The 4,000-point clouds are used to train the component in step 3, while the 1,000-point clouds are used to train the component in step 2, following the noise addition process described in \cite{ho2020denoising}. \\

\subsection{Shap-e}\label{sec:shape}
Shap-e \cite{Shap-E} is configured as a conditional generative model that, based on text or images, generates a three-dimensional representation. This development, conceived by OpenAI, distinguishes itself from Point-e by directly generating the parameters of an implicit function, which can be employed as a representation of a three-dimensional object, as required in NeRF (\ref{subsec:nerf}) and STF (\ref{subsec:stf}) representations. The model structure consists of two main components: an \textit{encoder} that produces the parameters of an implicit function from a representation of a three-dimensional object and a conditional diffusion model based on either images or text descriptions. \\

\begin{figure}[h]
    \centering
    \includegraphics[width = 0.49\textwidth]{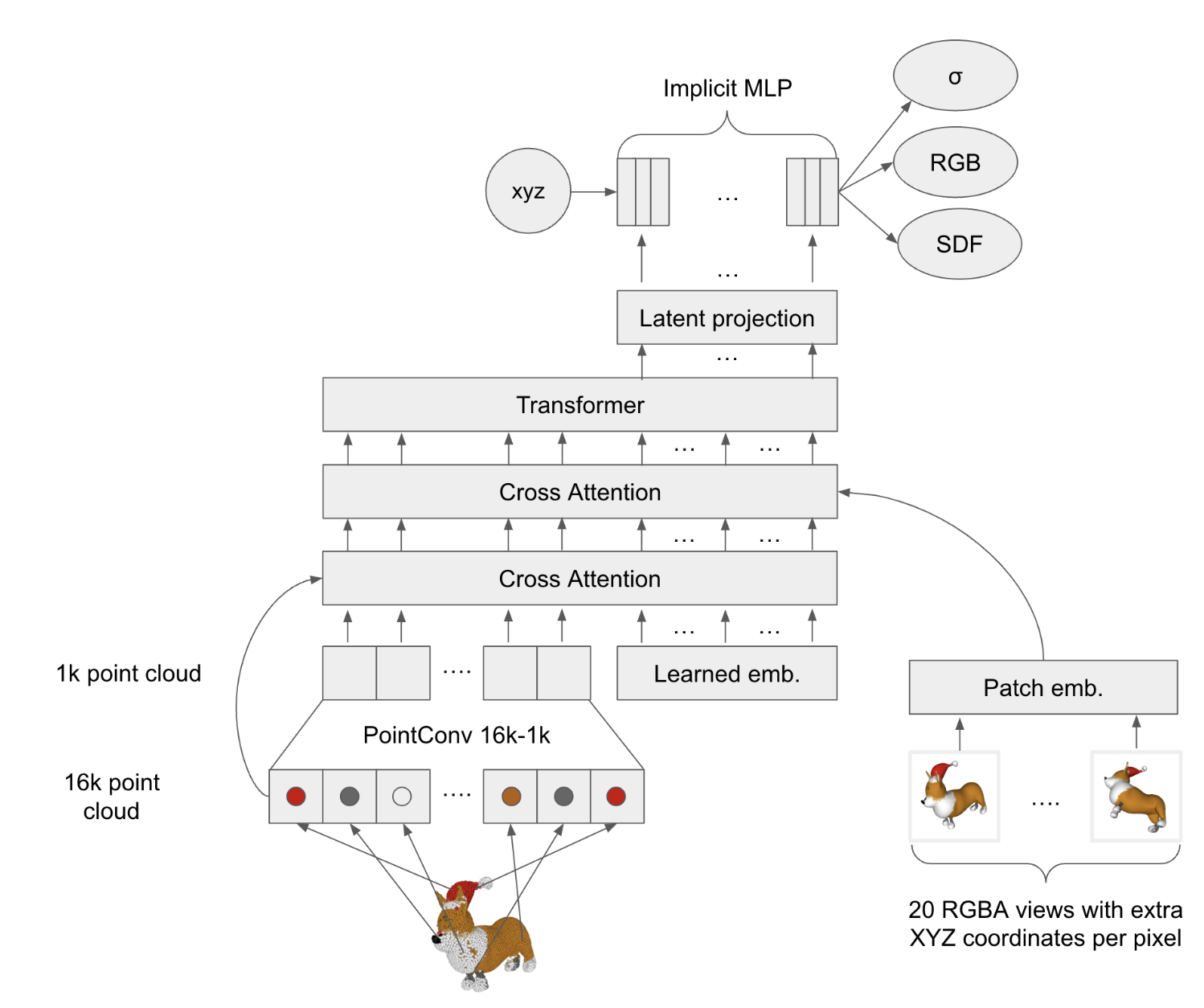}
    \caption{Encoder representation of Shap-e. Obtained from \cite{Shap-E}.}
    \label{fig:encoder}
\end{figure}

The encoder architecture is illustrated in Figure \ref{fig:encoder}. This component of the model takes as input two distinct representations of a three-dimensional object: a point cloud representation consisting of 16,000 points and a series of RGBA views of the object captured from various angles. These inputs undergo cross-attention layers, followed by a transformer-based architecture that generates a sequence of representations of the inputs. These representations are subsequently used as parameters in implicit functions that ultimately define the three-dimensional object. Specifically, the encoder output can be interpreted as the weights of an MLP neural network, similar to that used in NeRF, where, given an input $(x, y, z)$, three outputs are generated, corresponding to implicit function representations:
\begin{itemize}
    \item $\sigma$: Represents the density in the NeRF representation.
    \item $RGB$: Represents the color.
    \item $SDF$: Represents the signed distance, characteristic of the STF representation.
\end{itemize}

Initially, the encoder is trained exclusively to obtain NeRF representations. To achieve this, a combination of two loss functions is employed, using a joint function:
$$
\mathcal{L}_{NeRF} = \mathcal{L}_{RGB} + \mathcal{L}_T.
$$
The loss function $\mathcal{L}_{RGB}$ compares the colors of real 3D objects with those obtained from the implicit function generated by the encoder. More formally, a set $R$ of 4,096 rays (directions) is randomly sampled from the 3D object, and the estimated colors for each ray are compared with the implicit function and the actual color using the $L_1$ norm of the difference. The implicit function is evaluated through two rendering processes: a coarse rendering ($c$), which provides an initial less detailed approximation of the scene, and a fine rendering ($f$), which refines this initial estimation, producing a more detailed and precise representation of the scene. Using these renderings, the loss function is computed as follows:
$$\mathcal{L}_{RGB} = \mathbb{E}_{r \in R} \left[||\hat{C}_c(\textbf{r}) - C(\textbf{r})||_1 + ||\hat{C}_f(\textbf{r}) - C(\textbf{r})||_1 \right],$$
where $C(\textbf{r})$ represents the actual color, and $\hat{C}_c(\textbf{r})$ and $\hat{C}_f(\textbf{r})$ correspond to the predicted colors from the coarse and fine renderings, respectively.
\\

On the other hand, the loss function $\mathcal{L}_T$ compares the \textit{transmittance} \footnote{Transmittance: a property that describes how light travels through a medium or object. It represents the fraction of light that passes through a point along a ray without being absorbed or scattered. It is computed based on the densities $\sigma$.} obtained from the real object with that generated by the implicit function. Similar to $\mathcal{L}_{RGB}$, the comparison is based on the $L_1$ norm of the difference between the real and estimated values for the same set of rays in both renderings:
$$\mathcal{L}_{T} = \mathbb{E}_{r \in R} \left[||\hat{T}_c(\textbf{r}) - T(\textbf{r})||_1 + ||\hat{T}_f(\textbf{r}) - T(\textbf{r})||_1 \right].$$
Subsequently, the pretrained network is fine-tuned to incorporate the STF representation. For this purpose, a joint loss function is defined as:
\begin{align*}
    \mathcal{L}_{FT} = \mathcal{L}_{NeRF} + \mathcal{L}_{STF},
\end{align*}
where $\mathcal{L}_{STF}$ compares the actual mesh with the reconstructed one. To construct the mesh from the encoder outputs, a grid of dimensions $128 \times 128 \times 128$ is defined, and function values are computed over this grid. Using this information, $N$ images are generated via the rendering process, which are then compared with the real images using the $L_2$ norm to compute the final loss function $\mathcal{L}_{STF}$, following the formula:
$$\mathcal{L}_{STF} = \frac{1}{N\cdot s^2} \sum_{i=1}^N ||\text{Render}(Mesh_i) - \text{Image}_i||_2^2,$$
where $s$ is the resolution of the generated images, $Mesh_i$ represents the reconstructed mesh for sample $i$, and $\text{Image}_i$ is an RGBA target rendering for image $i$.\\

\begin{figure}[H]
    \centering
    \includegraphics[width = 0.5\textwidth]{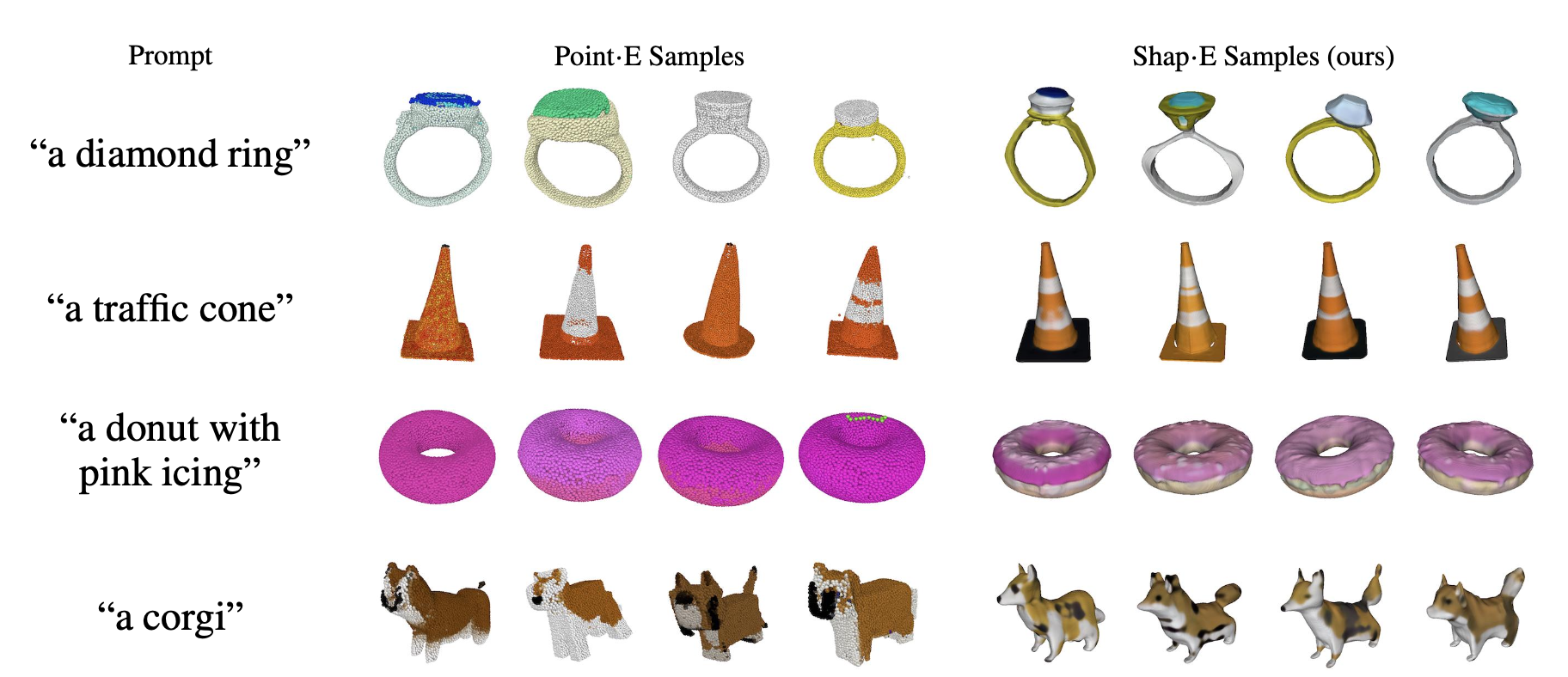}
    \caption{Comparison of Shap-e vs. Point-e results. Image obtained from  \cite{Shap-E}.}
    \label{fig:ejemplos}
\end{figure}

The second component of the model is a conditional diffusion model based on transformers, similar to Point-e (\ref{sec:pointe}). However, unlike Point-e, the input is not a point cloud but rather a sequence of latent vectors representing the weights of an implicit function. This sequence has dimensions $1024 \times 1024$, integrating it into the architecture as a sequence of 1024 tokens. For the conditional component of the diffusion model, the same approach as in Point-e (\ref{sec:pointe}) is adopted, utilizing CLIP embeddings for both text and images. Finally, unlike the original diffusion model, the objective is to minimize the error between \( x_0 \) and \( x_t \) rather than the noise error, as shown below:
$$
L_{x_0} = \mathbb{E}_{x_0 \sim q(x_0), \epsilon \sim \mathcal{N}(0, I), t \sim U[1, T]} \| x_\theta(x_t, t) - x_0 \|^2
$$
The results presented in the study include metrics evaluating the efficiency of the encoder, such as the PSNR metric and CLIP R-Precision, along with a qualitative analysis of the generated samples. The model demonstrates significant improvements over Point-e in the text-conditioned task, while in the image-conditioned task, it achieves comparable results. This is evident in the examples provided in the study, where the outputs of both models are compared (Figure \ref{fig:ejemplos}). Lastly, Shap-e exhibits faster inference than Point-e, as it does not require an additional diffusion model for upsampling, providing a notable advantage over the previously analyzed model.

\subsection{Alternative models}\label{sec:lrm} 

\subsubsection{Large Reconstruction Model}

The Large Reconstruction Model (LRM) \cite{hong_lrm_2023} introduces a data-driven approach for 3D reconstruction from a single image, utilizing a large-scale transformer-based encoder-decoder architecture \cite{vaswani_attention_2017}. This model consists of three main components:  

\begin{itemize}  
\item \textbf{Image Encoder:} LRM initially employs a Vision Transformer (ViT) \cite{dosovitskiy_image_2021}, specifically DINO \cite{caron_emerging_2021}, to encode the image into patch-wise feature tokens. DINO is chosen for its ability to self-supervise learning about the structure and texture of relevant image content, which is crucial for LRM to reconstruct both geometry and color in 3D space.  

\item \textbf{Image-to-Triplane Decoder:} A decoder is implemented to project image and camera features into learnable spatial-positional embeddings, translating them into triplane representations. This decoder acts as a prior network that provides essential geometric and appearance information, addressing the inherent ambiguities in single-image reconstruction.  

\item \textbf{3D Representation:} The triplane representation is adopted as a compact and expressive feature format. Each plane, $(64 \times 64) \times d_T$, is used to project any 3D point within the NeRF object's bounding box and retrieve point features via bilinear interpolation, which are subsequently decoded into color and density using an MLP. \\ 
\end{itemize}  

Finally, to address the Text-to-3D problem, a preliminary model would be required to generate the input image before applying LRM, such as Stable Diffusion \cite{rombach_high-resolution_2022}.\\

\subsubsection{Large Multiview Gaussian Model}

The Large Multiview Gaussian Model (LGM) is a framework designed to generate high-resolution 3D objects from a prompt or an image \cite{tang2024lgm}. From this work, two main ideas were identified: \\ 

\begin{itemize}  
\item \textbf{3D Object Representation:} The model leverages multiview Gaussian features as a mesh representation. This representation is subsequently fused for differentiable rendering.  
\item \textbf{3D Structure:} An asymmetric U-Net serves as the primary architecture for mesh generation. It operates on multiview images, which can be produced either from text or a single image using multiview diffusion models.  \\
\end{itemize}  

Multiview Gaussian models utilize a collection of 3D Gaussians to represent an object. Each Gaussian is defined by a center \( x \in \mathbb{R}^3 \), a scaling factor \( s \in \mathbb{R}^3 \), and a rotation quaternion \( q \in \mathbb{R}^4 \). Additionally, an opacity value \( \alpha \in \mathbb{R} \) and a color value \( c \in \mathbb{R}^C \) are included. These parameters define the Gaussian, which is rendered using alpha composition for each pixel.\\  

\begin{figure}[H]
    \centering
    \includegraphics[width = 0.5\textwidth]{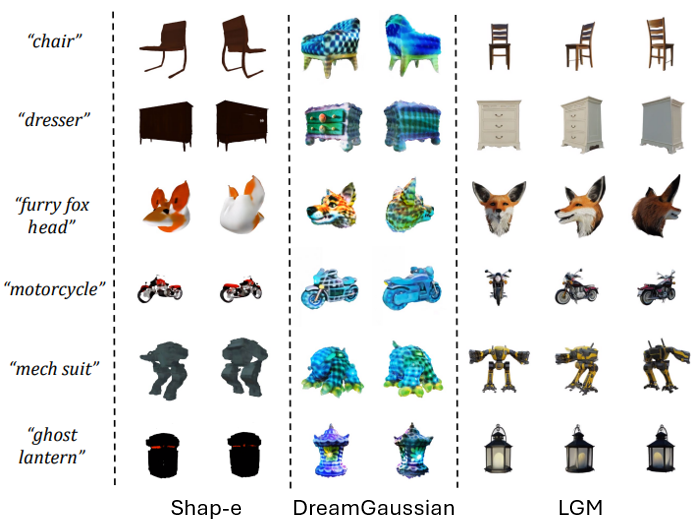}
    \caption{Comparison of results: LGM vs. Shap-e}
    \label{fig:LGM}
\end{figure}

In the asymmetric U-Net, each pixel extracted from the feature map is treated as a 3D Gaussian. The asymmetric U-Net allows the model to process higher-resolution images while controlling the number of Gaussians. In terms of performance, this model surpasses Shap-E and Point-E (see Figure \ref{fig:LGM}).

\section{Dataset}
\begin{figure}[h]
    \centering
    \includegraphics[width = 0.39\textwidth]{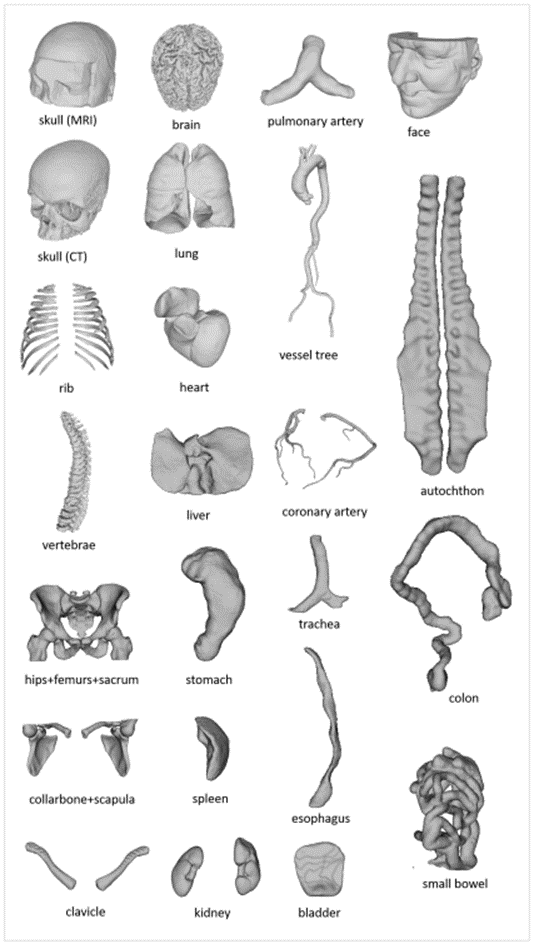}
    \caption{Example meshes in MedShapeNet, including various bones (skulls, ribs, and vertebrae), organs (brain, lung, heart, liver), vessels (aortic vessel tree and pulmonary artery), and muscles \cite{li2023medshapenet}. }
    \label{fig:ejemplos_data}
\end{figure}

To perform the fine-tuning phase of the model, the \textit{MedShapeNet} dataset was used as a source of biomedical meshes \cite{li2023medshapenet}. This dataset consolidates 23 different sources, comprising over 100,000 meshes with their respective annotations (see Figure \ref{fig:ejemplos_data}). Unlike existing datasets such as ShapeNet, which consist of computer-aided design (CAD) 3D models of real-world objects (such as airplanes, cars, chairs, and desks), MedShapeNet provides three-dimensional shapes extracted from patient imaging data, encompassing both healthy and pathological subjects.

The authors of MedShapeNet developed two interaction modes for utilizing this dataset: an internet-based interface that provides access to the original high-resolution shape data and an API that allows users to interact with mesh data through Python.  
\\  

The 3D figures are stored in standard formats for geometric data structures, namely NIfTI (.nii) for voxel grids, stereolithography (.stl) for meshes, and polygon file format (.ply) for point clouds, facilitating quick shape previews through existing software.  
\\  

\begin{figure}
    \centering
    \includegraphics[width = 0.45\textwidth]{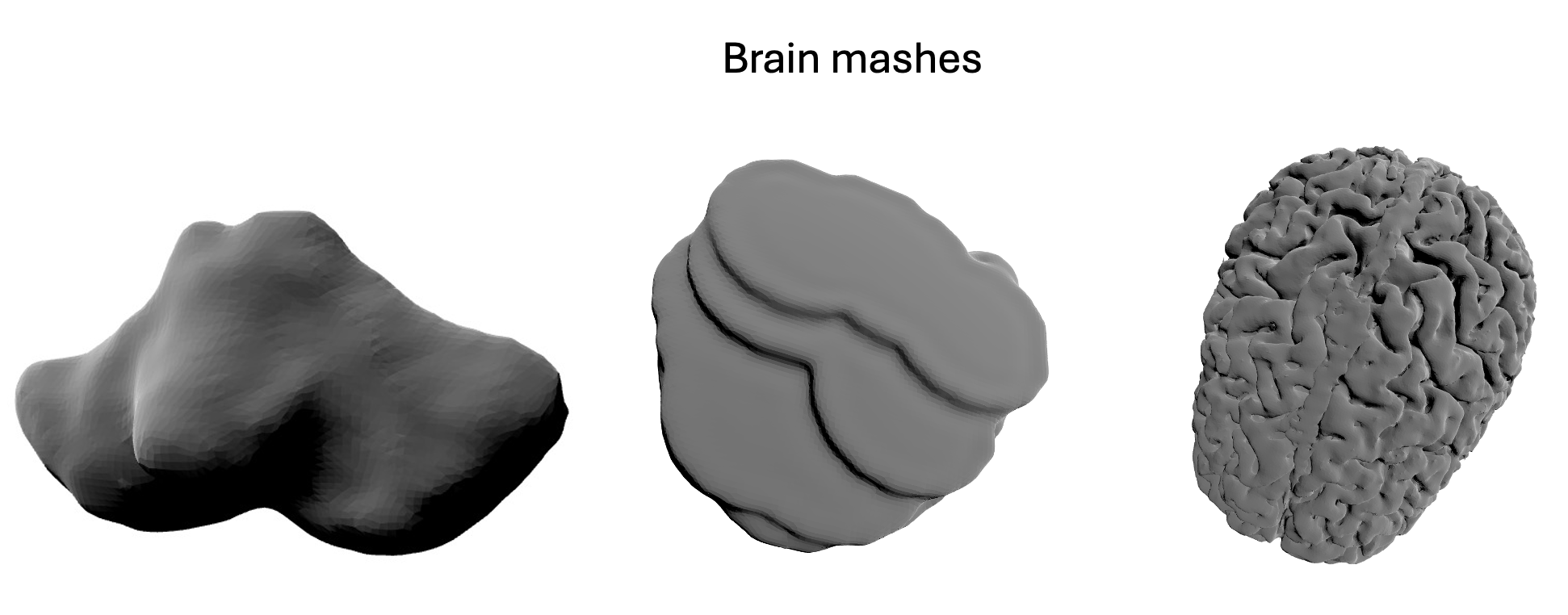}
    \caption{Examples of meshes from the brain category. On the left and center, two examples of misclassified objects. On the right, a correctly classified brain example.}
    \label{cerebro_feo}
\end{figure}

The dataset contains information on 50 different categories, of which 30\% correspond to organs, while the remaining 70\% comprise bones, muscles, arteries, veins, and various surgical instruments. Some categories of organs were found to exhibit issues with the provided meshes. In particular, such issues were observed in the \textit{brain} category, as illustrated in Figure \ref{cerebro_feo}, where several samples contained meshes that, at first glance, did not correspond to the assigned category. Another problematic category was \textit{heart}, where, although the general shape of a heart was evident, most samples displayed holes, possibly corresponding to adjacent arteries or veins.
\\

\begin{figure}[h!]
    \centering
    \includegraphics[width = 0.45\textwidth]{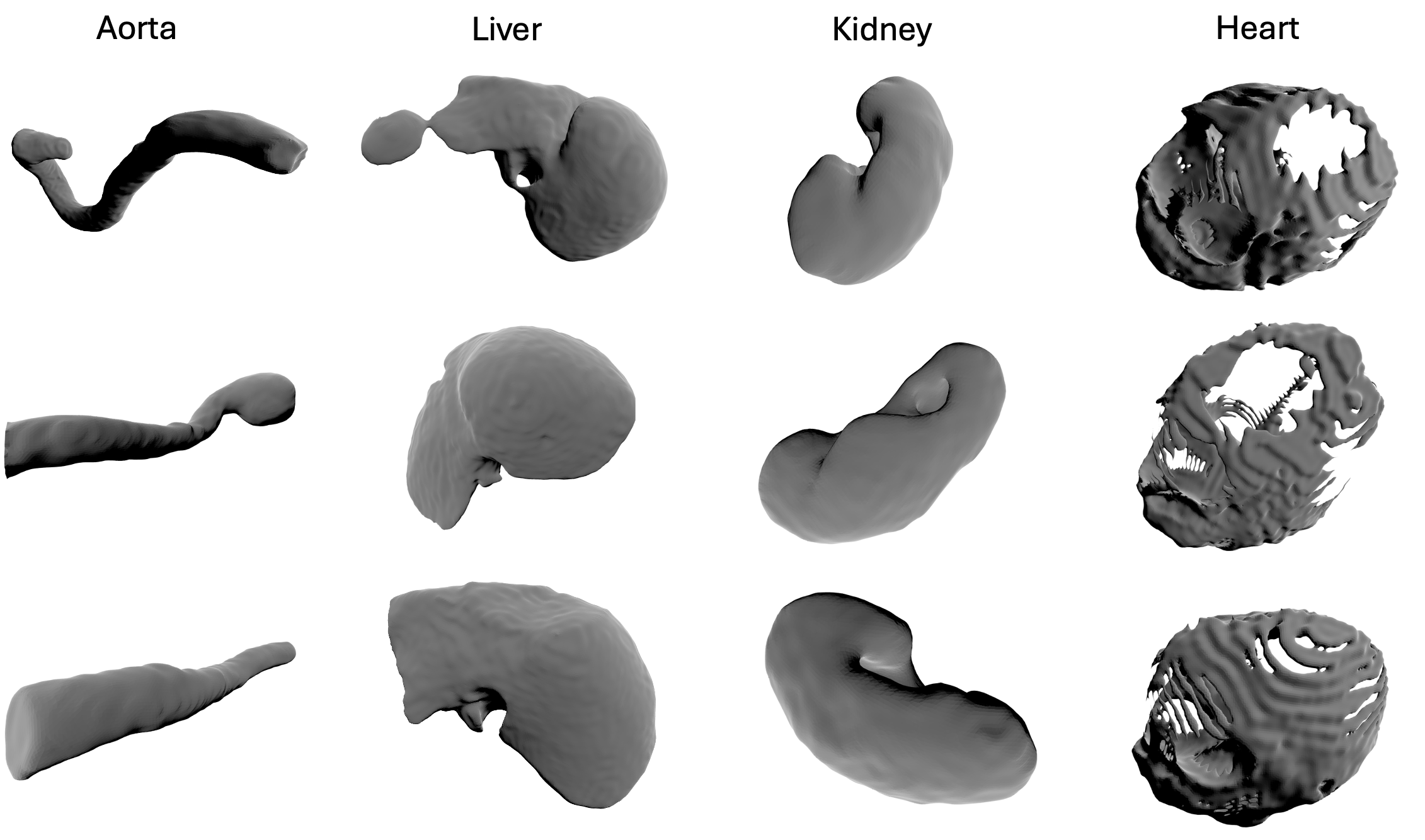}
    \caption{Examples of selected categories}
    \label{ejemplos_categorias}
\end{figure}

Due to the issues detected in certain categories and the limitation of computational resources, a total of 3,589 meshes were selected for fine-tuning. The selected meshes represent 10\% of the total organ data in MedShapeNet and belong to the categories of aorta, liver, kidney, and heart (Figure \ref{cerebro_feo}). Among these, the most represented category is \textit{Aorta}, the largest artery in the human body. Conversely, the least represented category in the selected training data is \textit{Heart}, with a total of 277 samples. As previously mentioned, this category contains instances with holes. By including it in the fine-tuning dataset, the importance of data quality and the complexity of the selected meshes in the model are assessed. Several examples from the selected categories are shown in Figure \ref{ejemplos_categorias}.

\begin{figure}[h!]
    \centering
    \includegraphics[width = 0.45\textwidth]{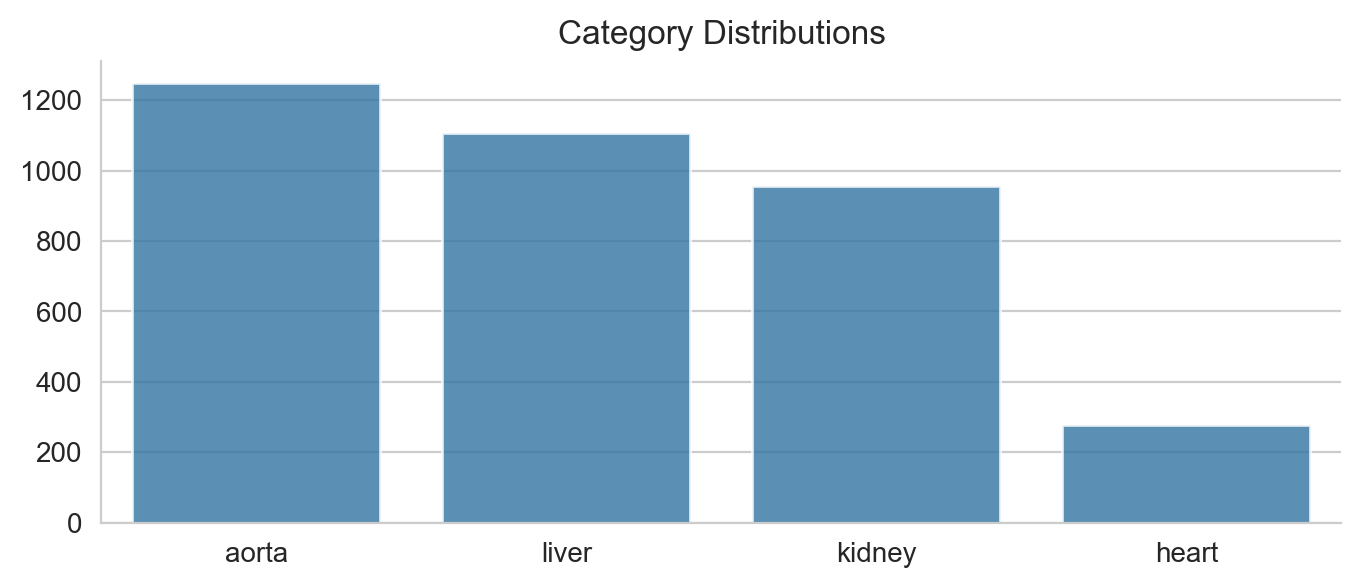}
    \caption{Category distribution}
    \label{distribucion_categorias}
\end{figure}

\section{Methodology}

In Section \ref{sec:relatedwork}, four candidate architectures were identified for use in the present biomedical problem. These four architectures are based on text/image encoding into a 3D representation and have open-source repositories, allowing for a detailed exploration of their functionality. Considering this, we followed the following experimental process:  

\begin{enumerate}  
    \item Selection of the model for fine-tuning  
    \item Data preprocessing  
    \item Fine-tuning parameters  
    \item Model evaluation
    \item Model deployment  
\end{enumerate}

\subsection{Selection of the model for fine-tuning }

We recognize that fine-tuning models trained on large generic datasets is always an ideal approach, whether in the context of text or image models. In this particular case, we aim to apply this technique to the domain of 3D objects, specifically biomedical objects. This allows us to leverage the prior knowledge of the pre-trained model, which has been trained on a vast array of objects. As previously mentioned, we conducted a study on existing models that convert text into 3D objects. However, we encountered issues with some of these models, which led us to select only one of them. \\  

Let us first examine LRM and LGM. These two models are the most recent within our group of text-to-3D object models. Both were trained on Objaverse \cite{objaverse}, a dataset containing more than 800,000 generic 3D objects along with their corresponding annotations (prompts). This makes them excellent candidates for leveraging their knowledge, not only due to the large amount of training data but also because they have demonstrated superior performance (see Figure \ref{fig:LGM}). However, these models are considerably large. LGM was trained using 32 NVIDIA A100 (80G) GPUs over four days, while LRM was trained using 128 NVIDIA A100 (40G) GPUs over three days. This immediately disqualifies them as viable options. \\  

The remaining options are Point-e and Shap-e. Both models were developed by OpenAI, but Shap-e is more recent. Additionally, Shap-e was trained on approximately one million more objects than Point-e. Furthermore, these models were trained using only 8 NVIDIA V100 GPUs, significantly fewer than LRM and LGM. For these reasons, we selected Shap-e as our model for fine-tuning. In the following sections, we will analyze the implications of selecting Shap-e as the model for our dataset.

\subsection{Data preprocessing}

The preprocessing pipeline is relatively simple and can be divided into two main steps: 
\begin{enumerate}
    \item \textbf{Conversion from STL to OBJ:}
    Since the MedShapeNet data was provided in the .stl format and the Shap-e model requires input in the .obj format, a format conversion was necessary before obtaining the latent representations of the meshes. Initially, the Aspose 3D API \cite{Apose} was used; however, due to a restriction limiting the conversion to a maximum of 50 meshes, we opted to use the open-source library Open3D \cite{open3d}.
    
    \item \textbf{Latent extraction:}  
    After obtaining the .obj files, we used Shap-e’s \textit{transmitter} (or encoder) to represent them in the latent space. This latent space corresponds to the weights of the MLP, which encode the object using NeRF, as discussed in Section \ref{sec:shape}. We did not fine-tune the \textit{transmitter}; that is, its weights were frozen, but we used it to convert our dataset into the latent representation. With these representations, we performed the fine-tuning of Shap-e's generative model.  
\end{enumerate}

\subsection{Fine-tuning parameters}

Fine-tuning, in general, follows certain best practices. Through fine-tuning, we aim to leverage a pre-trained model's knowledge to adapt it to our specific task, which, in this case, is mesh generation in the biomedical domain. To achieve this, we update the weights of the pre-trained model using our dataset. However, it is crucial to ensure that we do not degrade the model's existing knowledge, particularly in this case, the diffusion model. For this reason, we use a low learning rate of \(1 \times 10^{-5}\). Additionally, we employ a batch size of 8 and train for 25 epochs. The fine-tuning process is conducted on an NVIDIA A40-24C (24G) GPU.

\subsection{Model evaluation}

The fine-tuned model was evaluated using two approaches. In the first approach, we employ a quantitative evaluation, where the \textit{Mean Squared Error} (MSE) of the diffusion model serves as our evaluation metric. We compare this metric with the MSE of Shap-e without fine-tuning. To achieve this, we split the dataset into three subsets: 80\% for training, 10\% for evaluation, and the remaining 10\% for validation. It is important to highlight that this diffusion model operates in the latent representation of objects rather than on the objects themselves. This constitutes a major difference between Shap-e and Point-e. \\ 

Evaluating generative models poses a challenge due to their nature of generating novel content. This issue is also present in text and image generation models. For this reason, many evaluations rely on human assessors to determine the quality of the generated content. Consequently, our second evaluation method is qualitative. We compare our fine-tuned model against Shap-e, Point-e, LGM, and LRM by providing them with the same prompt within the biomedical domain. A visual comparison is then performed to determine which model yields the best results.

\subsection{Model deployment}

The deployment was carried out using Streamlit, a Python framework for deploying data science and machine learning applications. This framework operates as a state machine, where a series of stored elements are accessed from a front-end interface that can be viewed in a web browser. All components deployed using this framework function as a monolith, meaning that both the backend and frontend are hosted on the same machine, eliminating the need for additional components such as Docker or Triton. \\ 

To achieve this, an internal architecture was implemented with two subdivisions. The first subdivision corresponds to the backend, which is managed directly by the state machine. Here, the fine-tuned Shap-e model and its weights are stored, along with the prompt entered by the user for object generation. This state machine passes the prompt to the pretrained model to generate a prediction, which is then locally converted into an object file with a .ply extension using auxiliary functions from Shap-e.  \\

The second subdivision is the frontend, which is responsible for updating the state machine with user input, including the generation prompt and the selected model version. Once the user enters this information via the graphical interface in a web browser, the state machine updates with the specified model version and prompt. This update triggers the backend to load the corresponding model weights and use them to generate the .ply object associated with the user's prompt.

\section{Results and analysis}

\subsection{Learning during fine-tuning}
\begin{figure}[H]
    \centering
    \includegraphics[width = 0.4\textwidth]{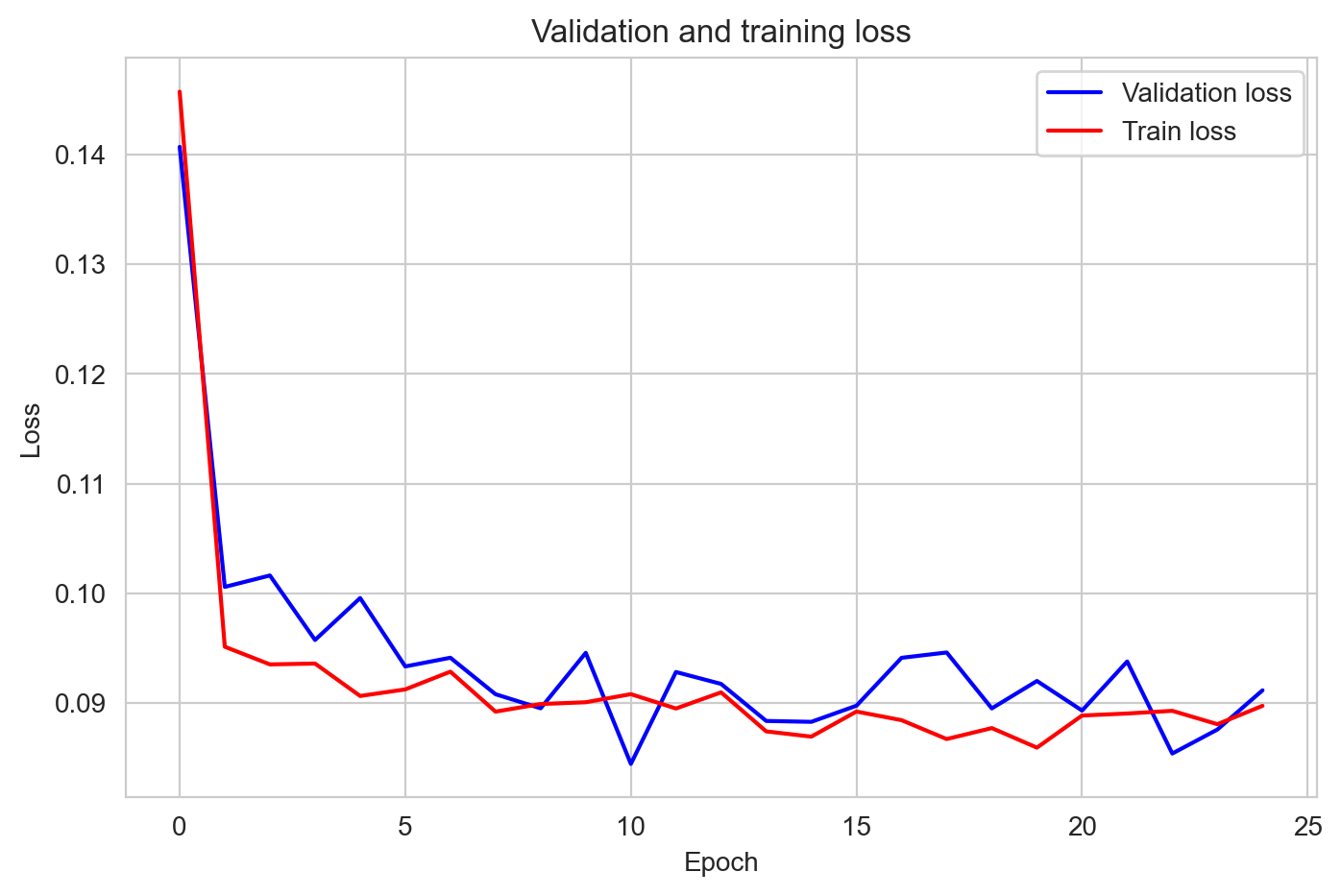}
    \caption{Mean Squared Error (MSE) loss function on training and validation data during fine-tuning.}
    \label{loss}
\end{figure}

During the 25 epochs of fine-tuning, a significant decrease in the Mean Squared Error (MSE) loss function was observed for both the training and test datasets (see Figure \ref{loss}). Initially, the loss function recorded values above 0.14 for both datasets, indicating a notable discrepancy between the model’s predictions and the ground truth values. However, as the training process progressed, a steady improvement was evident, with final loss values approaching 0.09. This decrease in the loss function reflects an enhanced accuracy and predictive capability of the model after fine-tuning on the pretrained data. \\

\begin{figure}[H]
    \centering
    \includegraphics[width = 0.45\textwidth]{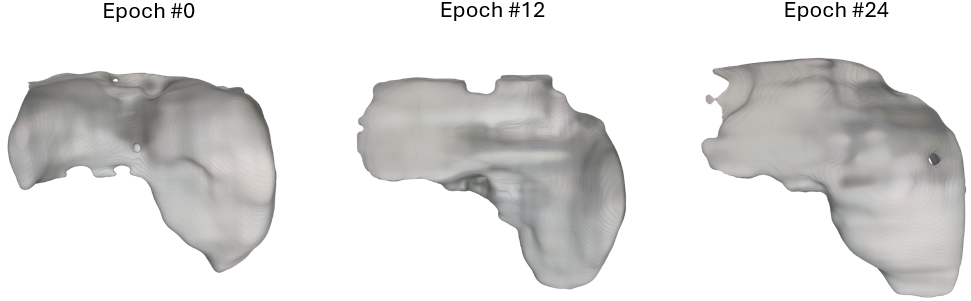}
    \caption{Evolution of the object (liver) throughout the training epochs.}
    \label{Evolution}
\end{figure}

Another way to visualize the model's learning process is by observing its improvement as the training epochs progress (see Figure \ref{Evolution}). Notably, it can be seen that the generated object (in this case, a liver) gradually improves its anatomical structure as the number of training epochs increases.

\subsection{Qualitative Comparison}

For the quantitative evaluation, we used the Mean Squared Error (MSE) metric on the evaluation dataset for our model and compared it with the MSE of Shap-e without fine-tuning. Table \ref{tab:MSE} shows the difference between the two models. It is evident that our model outperforms Shap-e, achieving an MSE lower by two orders of magnitude. This behavior is similar to what occurs in other generative models. Generally, well-established generative models are trained on generic datasets, such as language models and image models, making them effective for general tasks. However, when specialization in a specific task is required (e.g., generating texts of a particular literary genre for language models or realistic images of people for vision models), the best approach is to leverage the knowledge of the pre-trained model and fine-tune it with domain-specific data.  \\

We observe that this principle also applies to our case of biomedical object generation. However, a visual validation is still necessary. In Section \ref{ev_cualitativa}, we will determine whether this finding holds true through qualitative evaluation.

\begin{table}[h]
    \centering
    \renewcommand{\arraystretch}{1.5} 
    \setlength{\tabcolsep}{20pt}      
    \begin{tabular}{@{} l l @{}}
        \toprule
        \textbf{Model} & \textbf{MSE in latents} \\
        \midrule
        Shap-MeD & 0.089 \\
        Shap-e & 0.147 \\
        \bottomrule
    \end{tabular}
    \caption{Comparative table between our model and Shap-e, using the MSE of the latents in the evaluation dataset as the evaluation metric.}
    \label{tab:MSE}
\end{table}

\subsection{Qualitative Comparison} 
\label{ev_cualitativa}

\begin{figure}[H]
    \centering
    \includegraphics[width = 0.45\textwidth]{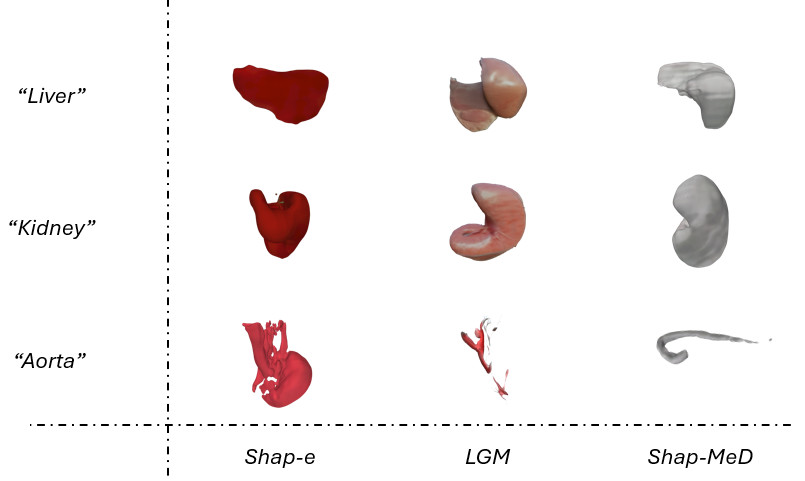}
    \caption{Comparison with Other Text-to-3D Models}
    \label{Comparacion}
\end{figure}

We compared our model with other text-to-3D models. In Figure \ref{Comparacion}, it can be observed that our model exhibits greater accuracy in the structure of the organs. Although LGM also produces good results, a closer analysis reveals that the liver lacks well-defined edges, reducing the precision of its structure. Additionally, the aorta generated by LGM lacks structural coherence, as it contains disconnected parts. \\  

Moreover, even more significant advantages can be seen when comparing our model with Shap-e. For instance, in Shap-e, the aorta is incorrectly represented as a heart with additional components, and the other organs lack the necessary precision to faithfully represent their real anatomical structure.

\section{Conclusion and Future Work}

The objective of this project was to develop a tool capable of transforming text into 3D objects specifically tailored for the biomedical field, aiming to streamline the modeling process. 3D modeling plays a crucial role in healthcare, as it is utilized in various applications, including surgical simulation and planning, personalized prosthetic implant design, dental implant development, medical education, and anatomical model creation.

To achieve this, we employed OpenAI's Shap-e, a generative text-to-3D model trained on over a million generic objects. We then fine-tuned it using biomedical-specific objects such as livers, aortas, kidneys, and hearts. On the evaluation set, our model achieved an MSE of 0.089, whereas Shap-e recorded an MSE of 0.147. Additionally, we conducted a qualitative evaluation, where our model outperformed larger state-of-the-art models.\\

Despite these promising results, this project still holds significant potential for improvement. First, it is crucial to consider both the quality and quantity of the dataset used. We observed that some objects were not rendered correctly, which negatively impacted the model's fine-tuning performance. Moreover, while our model surpasses larger models due to its specialization in our specific task, we believe that even better results could be achieved by fine-tuning one of these larger models, such as LGM or LRM. These models, in addition to being more complex, incorporate additional mechanisms that could enhance performance compared to Shap-e.

\printbibliography

\end{document}